\documentclass[%
 aps,
 superscriptaddress,
 amsmath,amssymb,
reprint,%
 numerical
]{revtex4-1}

\usepackage[utf8]{inputenc}
\usepackage[T1]{fontenc}

\usepackage{graphicx}

\usepackage{float}
\usepackage[dvipsnames]{xcolor}
\colorlet{green1}{green!40!black!100!}
\usepackage{dcolumn}
\usepackage{bm}
\usepackage{mathptmx}
\usepackage{etoolbox}
\usepackage{relsize}

\makeatother

\begin{document}

\preprint{APS/123-QED}

\title{Interaction between two polyelectrolytes in monovalent aqueous salt solutions}

\author{Xiang Yang}
\affiliation{Department of Applied Physics, Aalto University, P.O. Box 11000, FI-00076 Aalto, Finland}

\author{Alberto Scacchi}
\affiliation{Department of Applied Physics, Aalto University, P.O. Box 11000, FI-00076 Aalto, Finland}
\affiliation{Department of Chemistry and Materials Science, Aalto University, P.O. Box 16100, FI-00076 Aalto, Finland}
\affiliation{Academy of Finland Center of Excellence in Life-Inspired Hybrid Materials (LIBER), Aalto University, P.O. Box 16100, FI-00076 Aalto, Finland}

\author{Hossein Vahid}
\affiliation{Department of Applied Physics, Aalto University, P.O. Box 11000, FI-00076 Aalto, Finland}
\affiliation{Department of Chemistry and Materials Science, Aalto University, P.O. Box 16100, FI-00076 Aalto, Finland}
\affiliation{Academy of Finland Center of Excellence in Life-Inspired Hybrid Materials (LIBER), Aalto University, P.O. Box 16100, FI-00076 Aalto, Finland}

\author{Maria Sammalkorpi}
\affiliation{Department of Chemistry and Materials Science, Aalto University, P.O. Box 16100, FI-00076 Aalto, Finland}
\affiliation{Academy of Finland Center of Excellence in Life-Inspired Hybrid Materials (LIBER), Aalto University, P.O. Box 16100, FI-00076 Aalto, Finland}
\affiliation{Department of Bioproducts and Biosystems, Aalto University, P.O. Box 16100, FI-00076 Aalto, Finland}

\author{Tapio Ala-Nissila}\email{tapio.ala-nissila@aalto.fi}
\affiliation{Quantum Technology Finland Center of Excellence, Department of Applied Physics, Aalto University, P.O. Box 11000, FI-00076 Aalto, Finland}
\affiliation{Interdisciplinary Centre for Mathematical Modelling and Department of Mathematical Sciences, Loughborough University, Loughborough, Leicestershire LE11 3TU, UK}


\begin{abstract}
We use the recently developed soft-potential-enhanced Poisson-Boltzmann (SPB) theory to study the interaction between two parallel polyelectrolytes (PEs) in monovalent ionic solutions in the weak-coupling regime. The SPB theory is fitted to ion distributions from coarse-grained molecular dynamics (MD) simulations and benchmarked against all-atom MD modelling for poly(diallyldimethylammonium) (PDADMA).
We show that the SPB theory is able to accurately capture the interactions between two PEs at distances beyond the PE radius.
For PDADMA positional correlations between the charged groups lead to locally asymmetric PE charge and ion distributions. This gives rise to small deviations from the SPB prediction that appear as short-range oscillations in the potential of mean force. Our results suggest that the SPB theory can be an efficient way to model interactions in chemically specific complex PE systems.
\end{abstract}

\maketitle

\section{Introduction}

Polyelectrolytes (PEs) are polymers containing
electrolyte groups, which dissociate in aqueous solutions into solvated counterions and charged polymers. 
PEs exhibit physical properties very different from those of the uncharged polymers, in particular in terms of their aqueous solubility and tunability with salt~\cite{Barrat1997,Muthukumar2017}. 
Consequently, PE solutions and PE assemblies strongly react to changes in solvent environment, temperature, pH, and salt conditions~\cite{Stuart2010,DelasHerasAlarcon2005,Kocak2017}.
When their molecular persistence length is long enough, PEs can be considered, to some extent, as charged rods. Examples of such PEs include, e.g., functionalized cellulose fibrils, tobacco mosaic virus, DNA, and actin.
The interactions between rod-like PEs are widely addressed in chemistry and biology, leading to both rich functional and assembly behavior, such as in complex assembly of DNA and RNA~\cite{Wong2010,Seeman2017}, cellulose nanocrystal based advanced self-assembly~\cite{Habibi2010,Li2021}, and complexation based synthetic PE materials~\cite{VanderGucht2011,Stuart2010}.
The extensive applications of PEs in the field of nanostructured materials span from responsive materials~\cite{Stuart2010}, surface modification~\cite{Nemani2018} to layer-by-layer assembly~\cite{Tang1996,Sukhorukov1998}.

The fundamentals to understanding self-assembly, as well as structural and dynamical properties of dense PE systems, to a first approximation, come from the PE-PE pairwise interactions. These interactions in ionic solutions are however complex due to effects rising from both charge correlations and solution conditions.
Replacing atomistically detailed models with lower resolution, coarse-grained (CG) counterparts have paved the way to efficiently study and simulate large-scale processes at time scales inaccessible to all-atom models~\cite{Ingolfsson2014,Noid2013}.
To this end, it is beneficial to consider simplified geometries, such as assuming rod-like, rigid PEs, axially symmetric charge distributions, and simplified descriptions of ions in the solution around the charged species.

The mean-field Poisson-Boltzmann (PB) theory has proven to be efficient in describing the condensation of monovalent ions around single low-charge rods~\cite{Lamm1994,Naji2006,Deserno2000,Deserno2001,Robbins2014,Batys2017}.
This theory is able to predict various properties of PE solutions including electrophoretic migration, surface adsorption and osmotic pressure, for a wide range of concentrations~\cite{Menes1998,Takahashi1970,Schurr2003}.
However, PB theory fails when charge correlations
become strong, such as for high ion valency, high surface-charge density, and low temperatures.
There exists an extensive literature on the effects of charge correlations on ion condensation for PEs beyond the mean-field PB theory~\cite{Tan2005,Deserno2001a,Buyukdagli2014,Grochowski2008}.
Perhaps the most striking effect of such correlations is the reversal of the effective PE charge for multivalent counterions, which also reverses the field-driven PE mobility~\cite{Grosberg2002,Deserno2001a}.

Another shortcoming of these simplified models is that PEs, such as DNA or charged polypeptides, are atomistically structured, exhibiting a local spatially inhomogeneous charge distribution.
The details of the local structure are however important and affect both the persistence length~\cite{Tinland1997,Lu2002}, the charge distribution around the PE, and the PE interactions \cite{Antila2014,Antila2015}.
%
%
Ion condensation and salt solution response of PEs are also salt species and PE dependent~\cite{Antila2014,Antila2015}. Such dependencies can, to some degree, be
captured by an empirical modifications of the PB theory~\cite{Batys2017,Heyda2012}.

%




In systems consisting of two rod-like PEs, PB theory has also been often applied, in particular in the case of monovalent salt solutions.
The interaction between two charged cylinders has been calculated in many theoretical works with different boundary conditions and approximations~\cite{Harries1998,Overbeek1990,Hsu1999,Gilson1993}.
The linear version of PB theory with the Debye-H{\"u}ckel approximation (LPB) provides an explicit analytical solution of the interaction between two parallel cylinders~\cite{Ohshima1996,Brenner1973,Brenner1974}. 
However, LPB theory fails in dealing with highly charged molecules or small cylinder radii (as comparable to the Debye length), such as, e.g., in the case of DNA~\cite{Harries1998}.
Charge correlations in the strong-coupling regime lead to like-charge attraction~\cite{Buyukdagli2017,Buyukdagli2016b,Levin2002,Gronbech-Jensen1997,Arenzon1999,Naji2004,Butler2003}.
For dense PE systems, the attraction induces, e.g., bundle formation of F-actin and toroidal aggregates of concentrated DNA~\cite{Tang1996,Bloomfield1997}.

Recently, we have developed a soft-potential-enhanced PB (SPB) theory to efficiently
and accurately compute ion density distributions in the weak-coupling regime \cite{Vahid2022}. The soft potential in the SPB theory contains only one (physical) parameter, which can be fixed to give a good description for monovalent ion densities and electrostatic potentials around single rodlike PE for a wide range of salt and ion sizes. The SPB approach was shown to work well for monovalent salt concentrations up to 1 M and ion sizes ranging from those corresponding to small, hard monovalent ions, such as Na$^+$ and Cl$^-$, to almost an order of magnitude larger ions with diameter comparable to the PE diameter~\cite{Vahid2022}. 

Encouraged by the success of the SPB theory, in the present work we apply
it to compute the ion distributions and corresponding PE-PE interactions via the potential-of-mean-force calculation for two rodlike PEs in the weak-coupling regime. We compare the SPB modelling outcome against a chemically specific PE. To this end, we choose a relatively high-charge-per-length cationic and very common polyelectrolyte poly(diallyldimethylammonium) (PDADMA) as the PE focus of the study. PDADMA is chosen because of its technological relevance in, e.g., water purification, flocculation, and paper industry~\cite{Wandrey1999}.  Furthermore, the strong amide charges can be expected to cause localization, i.e. deviations from the mean field predictions. The SPB theory is benchmarked against molecular dynamics (MD) simulations of a CG model of PDADMA PE in salt solution, which allows the SPB fitting parameter to be optimized and fixed. We obtain good agreement between the
mean-field and coarse-grained models in capturing the
interactions between two PEs.
To further elaborate the influence of the atomic structure of the PE, we show
by atomistic-detail MD simulations of PDADMA how positional dependencies between the charged groups induce  correlation in the ion distribution. For the simulation setup in which the atomistic-detail PE is stretched straight, these emerge as asymmetric, helical configurations. Such positional correlations in the ion distribution lead to small deviations from the SPB and CG model
predictions in interaction strength and show as short-range deviation from the PB smooth prediction in the potential of the mean force.


The manuscript is organized as follows: Section II introduces the different methodologies used in this work. Specifically, in II-A we describe our all-atom MD simulations, in II-B the CG-MD simulations, and in II-C the SPB theory. In Section III-A we present the results regarding the SPB model parametrized against CG-MD simulations and compare the different models against the all-atom MD simulations. We discuss the ion density distributions predicted by the different approaches, underpinning the effect of the atomistic structure of the PE. Furthermore, in Section III-B we show good agreement between PE-PE interactions, as captured by the potential of mean force between two parallel PEs corresponding to PDADMA in monovalent salt solution obtained from the different description levels.
Finally, in Section IV we summarize our findings and provide prospects for the application of SPB to model chemically specific complex PE systems.

\section{Methods}\label{sec:methods}

\subsection{Atomistic molecular dynamics simulations}

\begin{figure}[h]
\centering
    \includegraphics[width=0.9\linewidth]{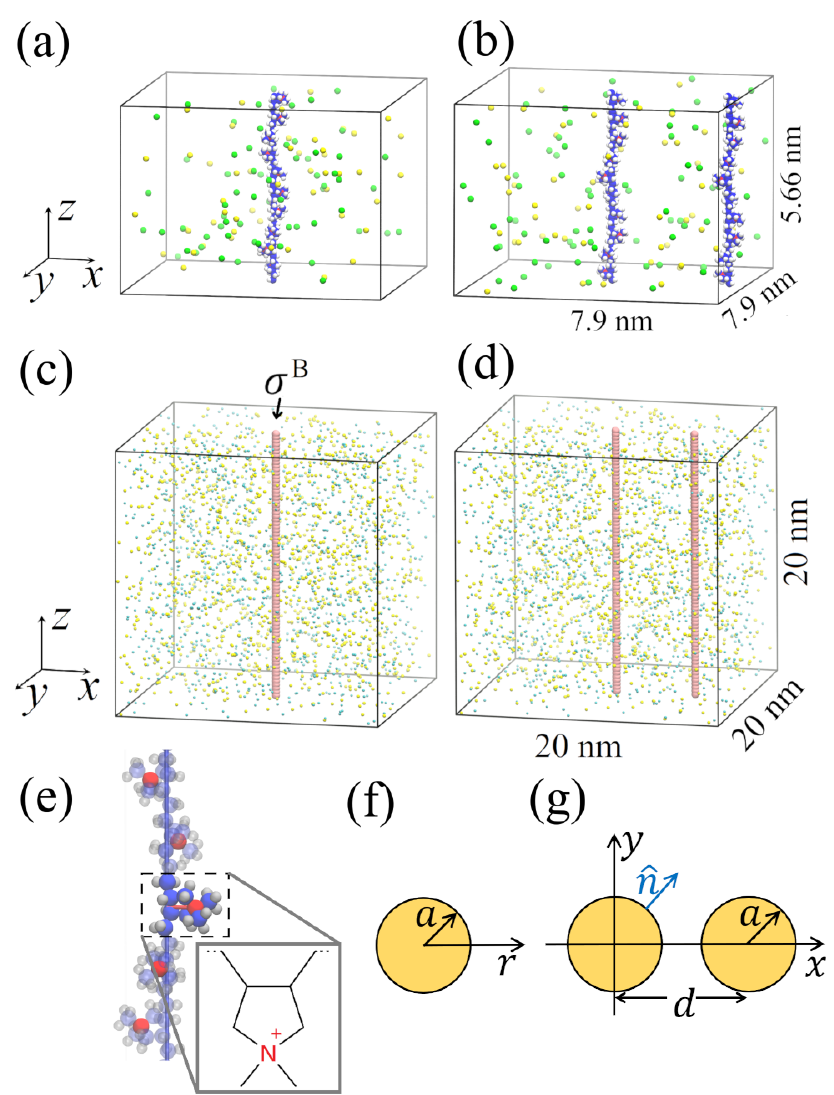}
    \caption{(a)-(b) Initial simulation snapshots of the all-atom MD simulations with (a) one and (b) two PDADMA chains in a $7.9\times7.9\times5.66$ nm$^{3}$ simulation box. For clarity, the explicit water in the simulations is omitted in these visualizations. (c)-(d) Initial simulation snapshots of the CG MD simulations of (c) one and (d) two CG models of PDADMA-like PEs in a $20\times20\times20$ nm$^{3}$ simulations box.
    $\sigma^{\rm B}$ is the effective bead diameter of PDADMA in the CG model, which has implicit solvent. In (a)-(d), the added salt concentration is $0.26$~M, and the Na$^+$ and Cl$^-$ ions are green and yellow, respectively. (e) A close-up view of the atomistic-detail structure of the straight PDADMA chain and the chemical structure of a monomer of PDADMA.
    In the atomistic structure, the N$^{+}$, C and H atoms (here spheres) are colored in red, blue, and gray, respectively. A $z$ axial straight line visualizes the PE axis.
    (f)-(g) Two-dimensional schematics of one and two cylinders in the PB theory with the relevant model variables (see text for details).
    \label{fig:system}}
\end{figure}

The system setup is shown in Figs.~\ref{fig:system}(a) and (b) where the atomistic-detail MD simulations include either one or two linear PE chains set to span the cuboid simulation box along the $z$ axis as periodic chains (terminal group connected over the periodic boundary condition). 
The initial configuration preparation protocol follows Ref.~\citenum{Batys2017} and leads to infinite, straight PE chains (covalently bonded over the periodic boundary).
The stretched PE configuration is used as a simplification, enabling direct comparison with PB theory for cylindrical objects. In comparison, a free-ended PDADMA chain fluctuates in backbone conformations, which influences both ion distribution and complexation with oppositely charged PEs, see e.g. Refs. \citenum{Batys2019,Batys2018,Yildirim2015,Diddens2019}.


PDADMA is chosen due to the relatively high charge per unit length and the  charged nitrogen residing away from the backbone axis which induces some spatial asymmetry.
The PE is fully charged (all monomers have dissociated electrolyte groups) and  Cl$^-$  ions are considered as the counterions to neutralize the polyelectrolyte charge. For excess salt, NaCl as dissociated ions is introduced.  
PDADMA is a relatively symmetric molecule in its charge distribution. However,
when set as an axially straight, periodic chain,
 the charged nitrogens (N$^{+}$) adopt a helical configuration around the PE main axis as their equilibrium configuration. This results from the charge groups repelling each other and steric effects. Additionally, methyl groups shield the nitrogens. PDADMA structure is shown in Fig.~\ref{fig:system}(e).

The atomistic simulations were performed using the GROMACS 2019.6 package~\cite{VanDerSpoel2005,Abraham2015}.
OPLS-aa force field~\cite{jorgensen1988} was used to describe the PE, whereas the explicit TIP4P model~\cite{Jorgensen1985} was employed for water molecules.
For the Na$^+$ and Cl$^-$ ions in the simulations the parameters originate from Refs.~\citenum{Aqvist1990} and \citenum{Chandrasekhar1984}, respectively.

The GROMACS solvate tool is used to solvate the PE chains. Excess salt (NaCl) in the concentration range $0.13 - 1$~M is introduced by replacing random water molecules by the ions. Finally, the equilibrated simulation box has dimensions $L_{x} = L_{y} = 7.9$ nm and $L_{z} = 5.66$~nm. The equilibrated straight PE chain length dictates $L_{z}$. 
For PDADMA, 10 monomers long chain is used based on
our earlier work~\cite{Batys2017} mapping this to be a sufficient length for the ion distribution to converge.


Van der Waals interactions are modelled using the Lennard-Jones potential. The particle-mesh Ewald (PME) method~\cite{Essmann1995} was applied for the long-range 
electrostatic interactions with a $0.16$~nm grid spacing and fourth-order spline.
Both the van der Waals and real-space electrostatics employ a $1.0$ nm direct cutoff (no shift).
All the bonds in the PE and water molecules were controlled by LINCS~\cite{Hess1997} and SETTLE~\cite{Miyamoto1992} algorithms, respectively. The stochastic $V$-rescale thermostat~\cite{Bussi2007} with a coupling constant of $0.1$~ps and reference temperature $T = 300$~K is used for temperature control.
On the other hand, pressure control is semi-isotropic with Parrinello–Rahman barostat~\cite{Parrinello1981,Nose1983} alongside a coupling constant of $1$~ps and reference pressure $1$~bar. Following Ref.~\citenum{Batys2017}, the system is set to be incompressible along the $z$ axis. A $2$ fs integration time step within the leap-frog scheme was applied in the $NpT$ simulations.

The initial configuration was first energy minimized via steepest descent algorithm 
until the largest force in the system was smaller than $1000$ kJ mol$^{-1}$nm$^{-1}$.
Next, a $2$ ns semi-isotropic $NpT$ simulation in which the positions of the heavy (i.e. non-hydrogen) PE atoms were kept fixed was performed to adjust the $x$ and the $y$ dimensions of the system as well as the distribution of water and ions around the PE. 
Finally, the PE positional constraints were released and a $100$ ns semi-isotropic $NpT$ simulation of a single PE was performed, of which the first 2 ns were disregarded from the analysis.


%

In order to examine the interaction between two PEs, a second PE was also placed parallel to the $z$ axis into an identical in size simulation box. The initial center-to-center distance of the $z$-axial PEs was set to be $d=3.5$~nm (cf. Fig.~\ref{fig:system}(b)). Solvation and initialization of the system followed the same procedure reported for the single-PE case.

An estimate of the potential of mean force (PMF) for the two PE interaction was obtained by umbrella sampling.
To generate the umbrella sampling configurations, the two PEs were pulled together at a fixed rate of $10$~nm/ns using a stiff spring ($k=3200\> {k}_{\textrm{B}}T/\textrm{nm}^{2}$). We sampled 21 configurations uniformly distributed at distances $d$ between $3.5 - 0.5$ ~nm. For each umbrella sampling window, a $20$~ns simulation with a harmonic tether potential with a spring constant of $k=320\> {k}_{\textrm{B}}T/\textrm{nm}^{2}$, but otherwise following the semi-isotropic $NpT$ setup described above, were performed.
The PMF was extracted using the Weighted Histogram Analysis Method (WHAM) of GROMACS~\cite{Kumar1992}.
The VMD package~\cite{Humphrey1996} was used for molecular visualizations.

\subsection{Coarse-grained molecular dynamics simulations}


Following Ref.~\citenum{Vahid2022}, the PE is modeled by a linear series of CG spherical beads. The CG PE is confined along the $z$ axis of a simulation box of size $20\times20\times20$ nm$^{3}$ with implicit solvent and periodic boundary conditions in all directions.
The initial configuration is shown in Fig.~\ref{fig:system}(c) using the VMD software package~\cite{Humphrey1996}.
Each bead carries a charge of $\zeta e$ where $e$ the elementary charge.
This is equivalent to a line charge density 
$\lambda=\zeta {e}/b$ where $b$ is the bead separation distance.
In total, $74$ consecutive beads, each with charge $\zeta e = 0.473 e$,  are set at a distance ${b}=0.27$ nm from one another.
This leads to a line charge density $\lambda = \zeta e/b = 1.75 $ e/nm, matching with the atomistic-detail PDADMA.
The ions were set initially at random positions in the simulation box but avoiding overlapping using the Packmol package~\cite{Martinez2009}.

The pair interactions between all the particles in the system are modelled via a standard Weeks-Chandlers-Andersen (WCA)~\cite{Weeks1971} potential of the form
\begin{equation}
       V^{ij}(r)= 4 \epsilon^{ij} \left[\left(\frac{\sigma^{ij}}{r}\right)^{12} -\left(\frac{\sigma^{ij}}{r}\right)^6 \right]+\epsilon^{ij};  \quad r<r^{ij}_{\rm c}.
\label{eq:WCA1}
\end{equation}
The labels $i,j$ denote either the CG ionic species (Na$^+$, Cl$^-$ CG equivalents) or the polymer beads (B), and
$r$ is the distance between the pair $ij$; $\sigma^{i}$ and $\epsilon^{i}$ denote the diameter and the depth of the potential well for species $i$.
We use the Lorentz-Berthelot (LB) mixing rule $\sigma^{ij}=(\sigma^{i}+\sigma^{j})/2$ and $\epsilon^{ij}=\sqrt{\epsilon^{i}\epsilon^{j}}$.
$r^{ij}_{\rm c}=2^{1/6}\sigma^{ij}$ is the cutoff radii of pair interactions.
We set $\epsilon^{\textrm{B}}=0.1$ kcal/mol, similar to the value of the ionic counterparts $\epsilon^{\textrm{Na}}=0.13$ kcal/mol and $\epsilon^{\textrm{Cl}}=0.124$ kcal/mol~\cite{Whitley2004,Siddique2013,Freeman2011}.
The diameter of the polymer beads $\sigma^{\textrm{B}}=0.54$ nm is determined by minimising the root-mean-squared error (RMSE) between $\rho^{\rm Cl}$ and $\rho_{\rm SPB}^{\rm Cl}$  (see Supplementary Material (SM)-1). The diameters of the CG ions $\sigma^{\textrm{Na}}=0.234$ nm and $\sigma^{\textrm{Cl}}=0.378$ nm are taken from Refs.~\citenum{Whitley2004,Freeman2011}.
The PE charge of $35 e$ is neutralized by $35$ monovalent counter-anions. In analogy to atomistic-detail simulations, we examine the system response to added monovalent salt concentrations of $0.13$ M, $0.26$ M, $0.52$ M, and $1$ M.

The electrostatic interactions were modeled via Coulombic potentials $\phi^{ij}_{\rm C}$.
The pairwise interaction between two ionic species $i$ and $j$ with charges $\zeta^{i}e$ and $\zeta^{j}e$ is given by
\begin{equation}
\beta e \phi^{ij}_{\rm C}(r)=\zeta^{i}\zeta^{j}\frac{l_{\rm B}}{r},
\end{equation}
where $\beta=1/k_{\textrm{B}}T$ and the Bjerrum length $l_{\rm B}=\beta {e}^{2}/(4\pi\varepsilon)$
is the distance at which two unit charges have interaction energy on the order of $k_{\textrm{B}}T$~\cite{Deserno2003,Deserno2000}. In water the permittivity is $\varepsilon=78\varepsilon_{0}$~\cite{Lide2004}, where $\varepsilon_{0}$ is the vacuum permittivity.
The implicit solvent is modeled by a continuum dielectric medium.

All CG MD simulations employ the LAMMPS Jan2020 package~\cite{Plimpton1995,Thompson2022}.
The long-range electrostatic interactions were calculated using the particle-particle particle-mesh method (PPPM)~\cite{Hockney2021}.
Coulombic pairwise interactions
were calculated in real space up to a cut-off of $r_{\rm c}^{i}=13\sigma^{i}$, where $i$ is any charged species, beyond which the contributions were obtained in reciprocal space.
All simulations were performed in the $NVT$ ensemble with a PPPM
accuracy of $10^{-5}$. The reference temperature was set to 300K, controlled by the Nose-Hoover thermostat~\cite{Nose1984,Hoover1985}.

Once the initial configuration was prepared, we performed an energy minimization of the system. After this, a 0.2 ns $NVT$ simulation in which the equations of motion were integrated with the velocity Verlet algorithm and 1 fs time step was carried out as initial equilibration of the system.
Finally, a 20 ns $NVT$ simulation was run for data analysis, out of which the first $1$ ns is disregarded. Here, a 2 fs time step was used. 

In the two-PE case, both PEs are also parallel to the $z$ axis. The first PE is set at the center of the box with $(x,y)=(0,0)$, while the second PE is set at $(x,y)=(d,0)$, as shown in Fig.~\ref{fig:system}(d).
With the exception of 70 CG Cl ions for neutralization, all initial settings for the two-PEs case are the same as in the single-PE case.

The equilibration and production run are performed analogous to the single PE case.
For determining the PMF between the two CG PEs, 15 two-PE configurations with PE-PE axial separation distance $d$ between 1 nm and 5 nm at equal intervals were generated. The PEs were kept fixed in position and for each fixed separation distance an MD run of 20 ns was performed.
The PMF vs distance $d$ was calculated by integrating the mean force $f_{\rm CG}$ over the separation distance as $\int_{d}^{\infty}f_{\rm CG}(x')dx'$.

\subsection{Soft-Potential-Enhanced Poisson-Boltzmann theory}

\begin{figure*}
\begin{centering}
\includegraphics[width=1.0\linewidth]{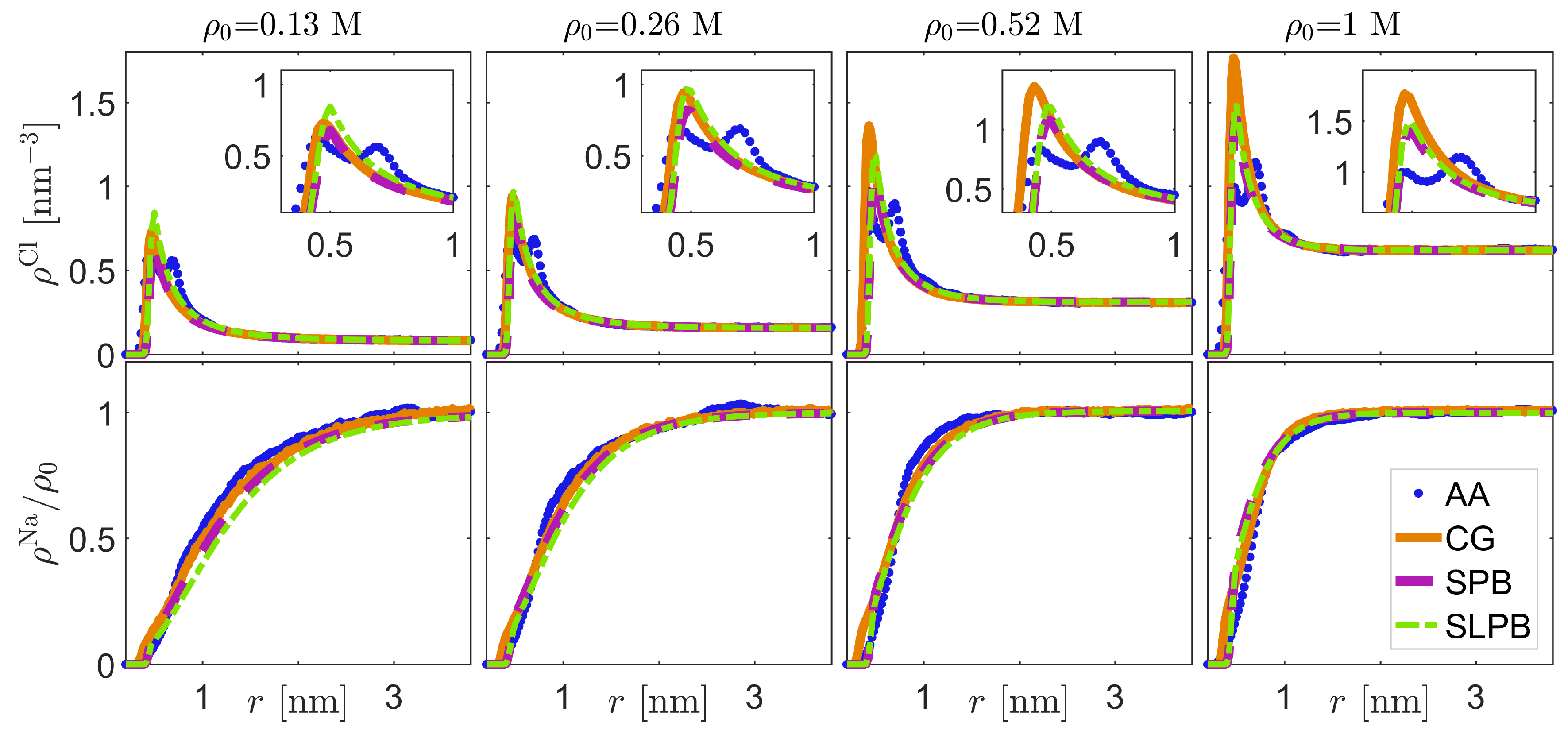}
\par\end{centering}
\caption{The radial number density profiles $\rho^{\rm Cl}(r)$ and $\rho^{\rm Na}(r)/\rho_0$ from all-atom (AA) simulations (large dots), CG model (solid lines), SPB equation (dashed lines) and SLPB equation (dash-dotted lines). The radial number density is calculated in the $xy$-plane as the function of distance $r$ measured from the PE backbone center axis (AA and CG simulations) or the rod center axis (SPB and SLPB models).
The insets show details of the data close to the peaks. Here
$\rho_{0}=$ 0.13 M, 0.26 M, 0.52 M and 1.0 M (from left to right).\label{fig:density_PDA1}}
\end{figure*}

Because of its success in describing monovalent ionic environments, the Poisson-Boltzmann (PB) equation is often used to describe the ion distribution and electrostatic potential in electrolyte solutions.
The PE is modelled as an impenetrable and rigid cylinder with surface charge density $\lambda/2\pi a$, where $a$ is the cylinder radius.
We consider here the case of positively charged PEs which have to be neutralized by adding negative counterions with number density $\rho^{\rm ci}$.
After this, salt is added to the system.
Thus, the number density of cations ($\rho_0^{\rm Na}$) equals that of the added salt $\rho_0$. The number density of anions ($\rho_0^{\rm Cl})$ then equals $\rho_0 + \rho^{\rm ci}$.
The full Poisson-Boltzmann (PB) equation for the electrostatic potential $\phi_{\rm PB}(r)$ surrounding such a cylinder can be written as
\begin{equation}
 \nabla^2 \phi_{\rm PB}(r)= -\frac{e}{\varepsilon} \sum_{ i} \zeta^i \rho_0^i \exp\Big(-\beta e \zeta^{ i} \phi_{\rm PB}(r)\Big),
\label{eq:PB}
\end{equation}
%
where $i= {\rm Na}$, ${\rm Cl}$, $\zeta^{i}$ is the ion valency and $\rho_0^{ i}$ the number density of the $i^{\rm th}$ ion species. The ion number densities from the PB theory can be obtained from
\begin{equation}
\rho_{\rm PB}^{ i}(r)=\rho_0^{ i} \exp\Big(-\beta e \zeta^{ i} \phi_{\rm PB}(r)\Big).
\label{eq:density_PB}
\end{equation}
In some cases we will also use data obtained from the linearized version of the Poisson-Boltzmann (LPB) equation, for which a simple closed-form analytic solution exists for two cylindrical polymers as explained in SM-2.

In a recent work~\cite{Vahid2022}, the prediction of ion number concentration from PB theory was enhanced with a cylindrical soft-potential correction (SPB) of the form
\begin{equation}
\rho^{i}_{\rm SPB}(r)=\rho_0^{i} \exp\Big({-\beta\zeta^{i} e \phi_{\rm PB}(r)-\beta\tilde{V}^{i}_{\rm }(r)}\Big),
\label{eq:density_SPB}
\end{equation}
which replaces the impenetrable cylinder in the standard PB theory with a soft cylinder potential $\tilde{V}^{i}_{\rm}(r)$. The effective potential $\tilde{V}^{i}_{\rm}(r)$ felt by the ions $i$ is defined as a cylindrically symmetric WCA potential, i.e
\begin{equation}
       \tilde{V}^{i}_{\rm}(r)=\begin{cases}4\tilde\epsilon^{{\rm B}i} \left[\left[\frac{\tilde \sigma^{{\rm B}i}}{r^{i}}\right]^{12} -\left[\frac{\tilde \sigma^{{{\rm B}i}}}{r^{i}}\right]^6 \right]
     +\tilde\epsilon^{{{\rm B}i}}, \> r^{i}<\tilde r^{{\rm B}i}_{\rm c};\\
     0, \>{\rm otherwise,}\end{cases}
 \label{effect_WCA}
\end{equation}
where the cutoff radius is defined by $\tilde r^{{\rm B}i}_{\rm c}=2^{1/6}\tilde\sigma^{{\rm B}i}$. Similar to Ref.~\citenum{Vahid2022}, we use $\tilde \sigma^{{\rm B}i}= \sigma^{{\rm B}i}$, $\tilde\epsilon^{\rm BNa}=0.107$ kcal/mol and $\tilde\epsilon^{\rm B Cl}=0.11$ kcal/mol, respectively.

To obtain accurate results from the SPB theory, the corresponding cylinder radius $a_{\rm SPB}$ has to be adjusted in Eq.~(\ref{eq:density_SPB}) for the electrostatic potential $\phi_{\rm PB}(r)$.
This effective radius may change with system conditions, such as, e.g., ion concentration.
The optimization is done by comparing Cl$^-$ ion number densities from the CG models with number density of the SPB theory prediction (Eq.~(\ref{eq:density_SPB})). 
From different values of $a$, Eq.~(\ref{eq:PB}) allows us to calculate $\phi_{\rm PB}(r)$, followed by Eq.~(\ref{eq:density_SPB}) to calculate $\rho_{\rm SPB}^{\rm Cl}$.
We identify the optimal radii $a^*_{\rm SPB}$ giving rise to the optimal potentials $\phi_{\rm SPB}^*(r)$ in Eq.~(\ref{eq:density_SPB}) based on minimizing the RMSE between $\rho_{\rm CG}^{\rm Cl}$ and $\rho_{\rm SPB}^{\rm Cl}$.
Additional information regarding soft-potential enhanced linear PB (SLPB) theory is provided in SM-2.
The SPB optimized radii (as well as those from SLPB) at different salt concentrations are summarized in Table S1 of the SM.
It is interesting to note that for different salt concentrations, the radius of the SPB theory only changes marginally ($\approx \pm 13\%$).
We thus use the average value of $a^*_{\rm SPB}=0.44$ nm throughout.

\section{Results and discussion}
\begin{figure*}[ht]
\begin{centering}
\includegraphics[width=0.9\linewidth]{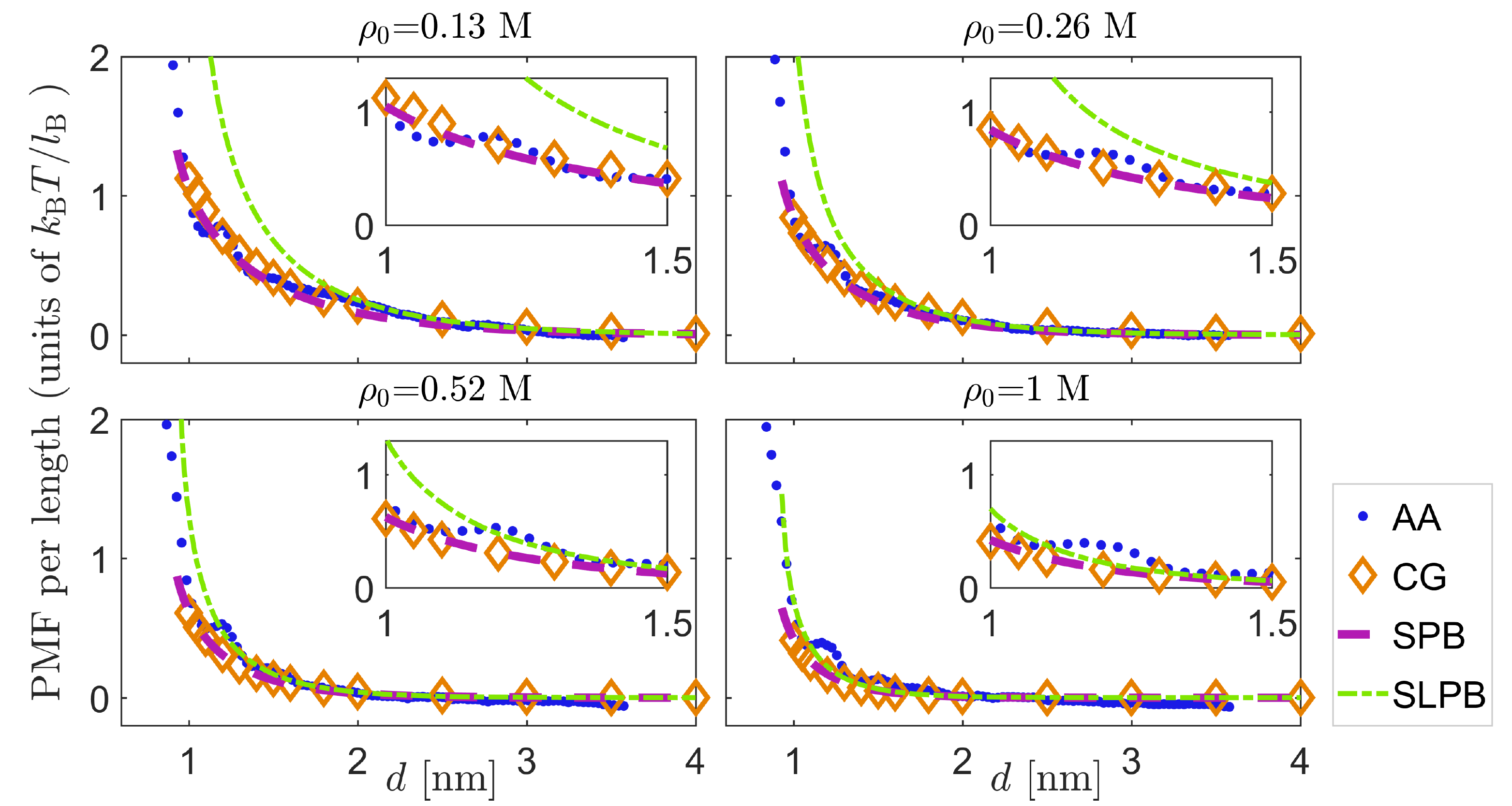}
\par\end{centering}
\caption{Comparison between the PMFs as a function of distance $d$ obtained from the different models and different salts. The abbreviations AA, CG, SPB, and SLPB denote all-atom MD, coarse-grained MD, soft-potential-enhanced Poisson-Boltzmann, and linearized SPB theories, respectively.
The insets show details of the oscillating regions in the PMF profiles.
\label{fig:PMF_PDA2}}
\end{figure*}
\subsection{Single-PE case}

We first assess the accuracy of the CG-MD simulations and the SPB and SLPB theories with respect to ion condensation around a single PE.
We start by setting up all-atom simulations of a single PDADMA at
added monovalent salt concentrations $\rho_{0}=$ 0.13 M, 0.26 M, 0.52 M and 1.0 M, 
respectively.
To calculate the radially symmetric (perpendicular to the $z$ axis) ion number density profiles $\rho^i(r)$, the $x$ and $y$ coordinates of the center of mass (CM) of the PE  have been taken as the reference point.
The density profiles $\rho^{\textrm{Na}}(r)$ and $\rho^{\textrm{Cl}}(r)$ at different salt concentrations are shown in Fig.~\ref{fig:density_PDA1}.
$\rho^{\textrm{Cl}}$ obtained from all-atom MD simulations exhibit two peaks at $r\approx 0.45$~nm and $r\approx0.7$ nm, respectively.
We note that the second peak increases as a function of salt concentration. These peaks are a result of the atomic level structure of PDADMA. Notably, the  charged N$^{+}$ that are usually found at roughly 0.25~nm from the center of the PE are responsible for the enhanced localization of the Cl ions. Details of the atomic configurations are presented in SM-3. To assess sampling equilibration, the orientational self-correlation function is calculated in SM-4. 

In the CG model and in SPB theory, we modelled PDADMA as a cylinder with effective radius $\sigma^{{\textrm{B}i}}$ and $a^*_{\rm SPB}$, respectively.
These are the only parameters to be optimized, and are obtained by minimizing the RMSE between the different models. Specifically, $\sigma^{\rm BCl}=0.46$ nm is obtained from averaging the minimised RMSE between all-atom MD and CG-MD simulations at different added salt concentrations (see the CG-MD simulations part of Methods).
On the other hand, the value of $a^*_{\rm SPB}=0.44$ nm is obtained by averaging the minimised RMSE between CG-MD simulations and SPB predictions at different added salt concentrations, as reported in the SM-1. Note that the latter value is also used for SLPB.
The comparison between SPB and SLPB theories is shown in SM-5.
In Fig.~\ref{fig:density_PDA1} we report the ion number density profiles from all the different models addressed here. We can see a satisfactory agreement between the different curves, with the exception of the small-scale oscillatory structure in the all-atom MD simulation results, mainly due to local asymmetry of PDADMA which cannot be captured by any of the CG techniques under the current assumptions.


\subsection{Two-PE case}

In the present case where the system has monovalent ions and is in the weak-coupling regime, there is no charge inversion and the two PDADMA molecules will always repel each other.
We quantified the interaction between two PEs via numerical computations of the PMF. Our results from all the different models and different salts here are summarized in Fig.~\ref{fig:PMF_PDA2}, with the all-atom MD being the ultimate benchmark for comparison. The first general observation is that the range of the PMF curves decreases with increasing salt concentration.
This is expected and is due to increasing screening effect mediated by the condensing anions. We find good agreement between most of the methods, with the exception of small-amplitude oscillations at $d\approx1.2$ nm that only show up in the all-atom MD data. This feature, as well as the differences rising from the different models, are addressed in more detail in the following sections.

\subsubsection*{The CG model and SPB theory}

In the case of the CG-MD model, the value $\sigma^{\textrm{B}}=0.54$ nm is inherited from the single-PDADMA case.
The CG model qualitatively captures the PMF curves when benchmarked against all-atom MD simulations at different salt concentrations up to 1~M (see Fig.~\ref{fig:PMF_PDA2}).
We find that due to the second PE, the condensation peak maximum of the Cl ions corresponds to $3-4$ times  higher number density than that of the single PE case at short distances. The detailed results on this are shown in Fig. S3 and S4 of the SM-6.

In order to obtain $\phi_{\rm SPB}$ (abbreviated with $\phi$ in this section), we employ a finite-element package (COMSOL5.2) 
to solve Eq.~(\ref{eq:PB}) in the $xy$ plane with the von Neumann boundary conditions at the surfaces of the two disks, i.e.  $\nabla\phi\cdot\hat{n}|_{x^2+y^2=a^2}=\lambda/2\pi a\varepsilon$ and $\nabla\phi\cdot\hat{n}|_{(x-d)^2+y^2=a^2}=\lambda/2\pi a\varepsilon$. To do so, we use $a=a^*_{\rm SPB}=0.44$ nm as obtained in the single-PE case. The same is applied to the SLPB theory. Here $\hat{n}$ is the unit normal vector of each surface, as sketched in Fig.~\ref{fig:system}(g).
The mean electrostatic force between the two cylinders per unit length $f(d)$ at distance $d$ was calculated via the stress integral~\cite{Harries1998}
\begin{equation}
\begin{split}
f(d)=\varepsilon\int_{0}^{\infty}\bigg[&2\left(\frac{\kappa}{\beta e}\right)^{2}\left(\textrm{cosh}(\beta e\phi)-1\right) \\& +\left(\frac{\partial\phi}{\partial y}\right)^{2}-\left(\frac{\partial\phi}{\partial x}\right)^{2}\bigg]_{x=d/2}dy,\label{eq:force_2PE}
\end{split}
\end{equation}
where $\kappa= \sqrt{e^2(\rho^{\rm Na}_0+\rho^{\rm Cl}_0)/\epsilon k_{\rm B}T}$ is the screening parameter. The mean force can be integrated to get the potential of mean force between two cylinders as $V_{\rm PMF}(d)=\int_{d}^{\infty}f(x')dx'$.

The SPB prediction of the PMF is in good agreement with the CG model and the all-atom MD simulations, as shown in Fig.~\ref{fig:PMF_PDA2}. In contrast, the PMFs from linear SPB are inaccurate at low salt concentrations.  The linear SPB predictions are discussed in more depth in SM-5.


\begin{figure}[ht]
\begin{centering}
\includegraphics[width=0.9\linewidth]{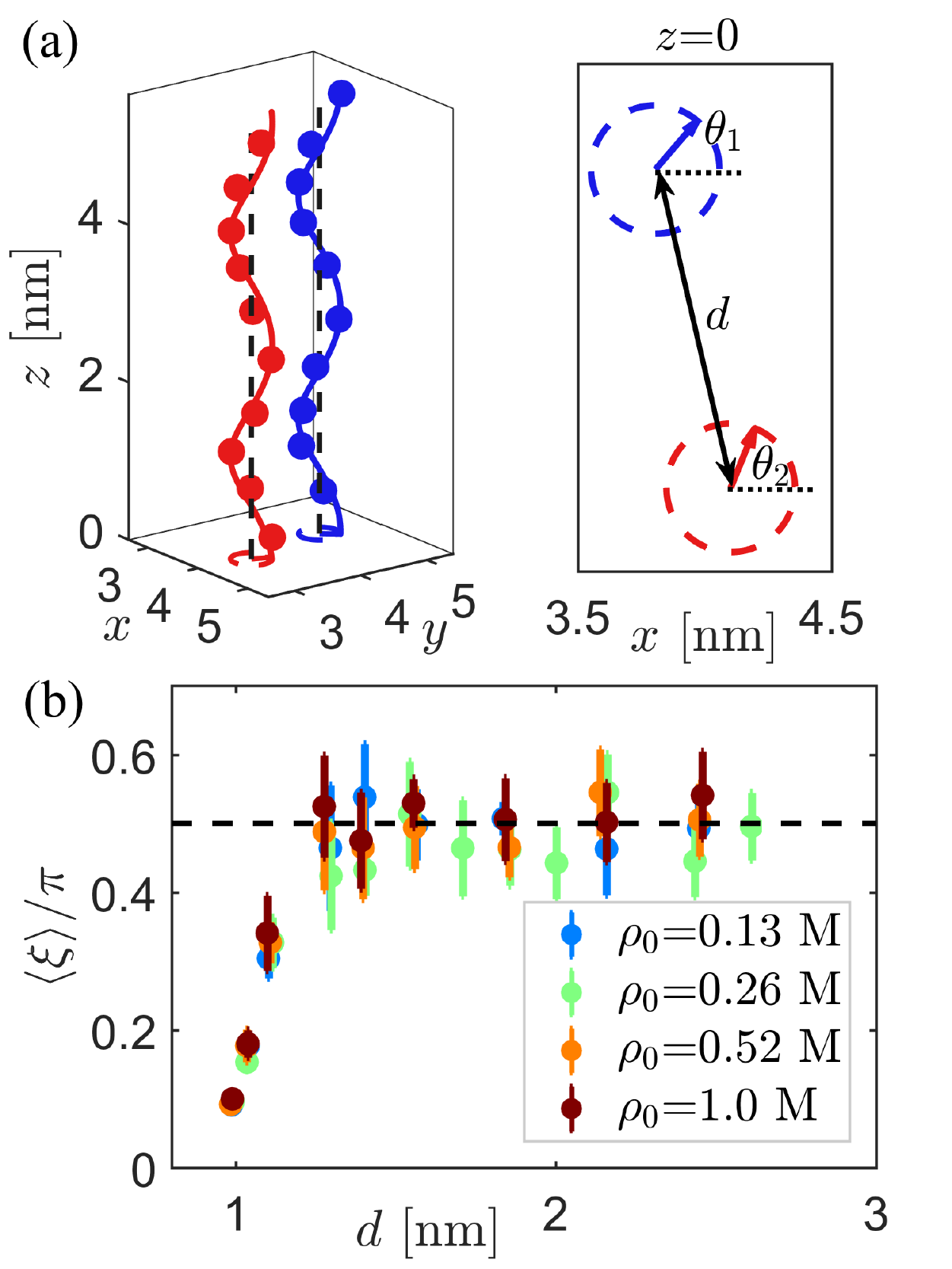}
\par\end{centering}
\caption{(a) A schematic representation of the charge group positioning in the two PDADMA chains on the left panel; the circles represent charged N$^{+}$ atoms. A cross-section at the $z=0$ plane is shown on the right panel.
The orientations of the helices with respect to the positive $x$ direction are defined as $\theta_{1}$ and $\theta_{2}$, and the relative orientation is $\xi\equiv\left|\theta_{1}-\theta_{2}\right|$.
(b) $\left\langle \xi\right\rangle /\pi$ as a function of the distance $d$.
The dashed line represents $\left\langle \xi\right\rangle /\pi=1/2$ which corresponds to uncorrelated orientations of the two molecules at long distances (see text for details).
\label{fig:phase_PDA2}}
\end{figure}

\subsubsection*{The influence of the atomistic structure}

When the two atomistic detail PDADMA chains in the all-atom MD simulations are separated by a distance comparable to the size of a monomer, the orientations of the PE side chains will affect their interaction.
As shown in Fig.~\ref{fig:system}(e), the 10 charged N$^{+}$ atoms of each PDADMA chain form in the elongated, axially straight configuration a right-handed cylindrical helix.
In cylindrical coordinates, with the longitudinal axis centered at the backbone of each PDADMA chain (see the two black dashed lines in left panel of Fig.~\ref{fig:phase_PDA2}(a)), the N$^+$ atom of the $l^{\rm th}$ monomer of the $k^{\rm th}$ PE has coordinates $(r^l_k, \theta^l_k, z^l_k)$, with $k=1,2$ and $l=1,2,..., 10$.
We take the positive $x$ direction to correspond to $\theta = 0$.
We set a plane at $z=0$, where the orientations of the two helices are defined by the angles $\theta_{k}$ ($k=1,2$), respectively, as shown in the right panel of Fig.~\ref{fig:phase_PDA2}(a).
Because the pitch of the helices is $L_{z}/2$, the angles $\theta_{k}$ can be calculated from the location of the first monomer (i.e. $l=1$) as $\theta_{k}= \theta^1_k- 4\pi z^1_k/L_z$.
We then define the variable $\xi\equiv \vert \theta_{1}-\theta_{2} \vert$ to represent the relative orientation between two PDADMA chains.
In Fig.~\ref{fig:phase_PDA2}(b) we show the average value of $\xi/\pi$ as a function of the distance $d$ between the two PEs.
At long distances, the positioning of the PDADMA charges remains uniformly random with $\left\langle \xi\right\rangle/\pi =1/2$.
In contrast, at small PE-PE separations of $d < 1$ nm, the charge group positioning is highly correlated which shows as two orientations coupling strongly. The coupling rises from minimization of electrostatic and steric repulsion by rotation and axial displacement of the PDADMA chains. Nevertheless, the outcome is a strong interchain repulsion evident in the corresponding PMFs.

\section{Summary and Conclusions}

In this work we have addressed the interactions between two PEs by using the recently developed soft-potential-enhanced Poisson-Boltzmann (SPB) theory.
We have here focused on the case of PDADMA molecules in monovalent salt solutions in the weak-coupling regime, where the interactions are expected to be fully repulsive at all distances.
To this end, we have first addressed the ion number density around a single PDADMA.
In the elongated, axially straight PE configuration, the charged N$^+$ atoms of PDADMA adopt a helical configuration as their minimum energy configuration by electrostatics and sterics. This leads to the two peaks in the ion distribution around the atomistically detailed PE (see Fig.~\ref{fig:density_PDA1}). Although the isotropic SPB theory cannot reproduce such features, it can qualitatively capture the atomistic ion distributions at salt concentrations up to 1~M.
By minimizing the RMSE between the number density distribution of the CG model which has been fitted on all-atom MD simulations and the one obtained from SPB theory, we find an optimal effective radius for PDADMA ($a^*_{\rm SPB}=0.44$ nm), which is the only (physical) fitting parameter in the SPB theory.

To study the interactions between two PEs we have considered two parallel rodlike PDADMA chains and numerically computed the PMF curves at different salt concentrations.
We find that the predictions from the SPB method are accurate when the distance between the two PDADMAs is larger than their radius. There are small oscillations in the PMFs  at short distances as revealed by all-atom MD simulations. These oscillations, rising from the chemical structure of the PE, are due to the coupling between charged atoms, steric packing of the side chains, and correlations in the relative orientations of the chains at distances smaller than about $1.3$~nm.

While our work suggests that the SPB theory is a simple and accurate way to model interactions in complex PE systems for a large range of system parameters, the limitations of the theory should be kept in mind.
We naturally expect the theory to fail when the approximations in the weak-coupling theory are no longer valid. This would happen in cases where the PE line charge density is high, there are electrostatic correlations arising from strong Coulombic interactions, or the assumption of a straightened PE as in our simulation setup leads to a large entropy loss that destabilizes such configurations (see SM-7 for some additional discussion).

\section*{Acknowledgements}

This work was supported by Academy of Finland grant Nos. 307806 (T.A-N.) and 309324 (M.S.) and the Academy of Finland Center of Excellence Program (2022-2029) in Life-Inspired Hybrid Materials (LIBER), project number (346111) (M.S.), as well as a Technology Industries of Finland Centennial Foundation TT2020 grant (T.A-N.). We are grateful for the support by FinnCERES Materials Bioeconomy Ecosystem. Computational resources by CSC IT Centre for Finland and RAMI -- RawMatters Finland Infrastructure are also gratefully acknowledged.

\bibliography{reference0420_abbr}
\bibliographystyle{ieeetr}

\pagebreak
\clearpage

\onecolumngrid
\begin{center}
{\Large \bf Supplementary Material}
\end{center}
\vspace{8mm}
\twocolumngrid

\setcounter{equation}{0}
\setcounter{table}{0}
\setcounter{figure}{0}

\renewcommand{\theequation}{S\arabic{equation}}
\let\oldthetable\thetable
\renewcommand{\thetable}{S\oldthetable}
\renewcommand{\thefigure}{S\arabic{figure}}

\section*{1. Root-mean-square error analysis}\label{sec:RMSE}

We optimize the parameters $\sigma^\textrm{B}$ and $a^*_{\rm SPB}$ by minimizing the root-mean-square error (RMSE) between the Cl$^-$ density profile obtained from the different models.
The definition of RMSE between two density profiles $\rho_1(r)$ and $\rho_2(r)$ is defined as:
\begin{equation}
{\rm RMSE} =\frac{1}{\sqrt{N}}\sqrt{\sum_{k=1}^N \mid \rho_{1}(r_k)-\rho_{2}(r_k)\mid^2},
\label{eq:RMSE}
\end{equation}
where $N$ is the number of sampling points.

Specifically, in order to obtain the effective diameter of the PE in the CG model, $\sigma^{\rm B}$, $\rho_1$ is replaced by $\rho^\textrm{Cl}$ from all-atom MD simulations, whereas $\rho_2$ by $\rho^\textrm{Cl}_{\rm CG}$, obtained from the CG model, which depends on $\sigma^{\rm B}$.
The values of the optimized diameter $\sigma^{\textrm{B}}$ for different salt concentrations are shown in Table.~\ref{tab:RMSE}.
We choose the average over the four salt concentrations, i.e. $\sigma^{\textrm{B}}=0.54$ nm.

On the other hand, in the case of SPB theory, $\rho_2$ is replaced by $\rho^{\rm Cl}_{\rm SPB}$, which depends on $a_{\rm SPB}$.
The optimal values from fitting $a^*_{\rm SPB}$ at different salt concentrations are shown in Table \ref{tab:RMSE}.
Also in this case we choose the average over the different salt concentrations, i.e. $a^*_{\rm SPB}=0.44$ nm, used throughout the work (also for the SLPB theory).

\begin{table}[hb]
\small
\caption{The optimal values of $\sigma^{\textrm{B}}$, $\sigma^{\textrm{BCl}}$ obtained for the CG model and $a^*_{\rm SPB}$ for the SPB theory in different salt concentrations.
\label{tab:RMSE}}
\begin{tabular*}{0.48\textwidth}{@{\extracolsep{\fill}}lllll}
\hline
$\rho_{0}$ [M] & 0.13 & 0.26 & 0.52 & 1.0\tabularnewline
\hline
$\sigma^{\textrm{B}}$ [nm] & 0.54 & 0.54 & 0.52 & 0.54\tabularnewline
$\sigma^{\textrm{BCl}}$ [nm] & 0.46 & 0.46 & 0.45 & 0.46\tabularnewline
$a^*_{\rm SPB}$ [nm] & 0.50 & 0.45 & 0.43 & 0.39\tabularnewline
\hline
\end{tabular*}
\end{table}

\section*{2. Soft-potential-enhanced linear Poisson-Boltzmann theory}\label{sec:SLPB}

When $\beta e\phi(r)\ll1$, which corresponds to small electrostatic potentials, we can linearize the full PB equation. This linear approximation is commonly referred to as the Debye-H\"uckel approximation. Equation~(3) in the main text then becomes
\begin{equation}
 \nabla^2 \phi_{\rm LPB}(r)=\kappa^2 \phi_{\rm LPB}(r),
 \label{eq:LPB}
\end{equation}
where $1/\kappa$ is the Debye or screening length~\cite{Grochowski2008}.
The analytic solution of Eq.~(\ref{eq:LPB}) for $r>a_{\rm }$ is~\cite{Brenner1974}
\begin{equation}
\phi_{\rm LPB}(r) =\frac{\lambda}{2\pi a\kappa \varepsilon }\frac{K_{0}(\kappa r)}{K_{1}(\kappa a_{\rm })},
\label{eq:sol_LPB}
\end{equation}
where $K_0$ and $K_1$ are modified Bessel functions of order zero and one, correspondingly.
The soft-potential enhancement can be used for both the full and linearized PB theories.
Finally, for the two-PE case, the corresponding force from the linear theory $f_{\rm SLPB}$ and the PMF are obtained by numerically solving Eq. (\ref{eq:LPB}) similarly to the full PB equation, with the boundary conditions at the surface of the two disks with radii $a=a_{\rm SLPB}$ (cf.  Fig.~1(g) in the main text) $\nabla\phi\cdot\hat{n}|_{x^2+y^2=a^2}=\lambda/2\pi a\varepsilon$ and $\nabla\phi\cdot\hat{n}|_{(x-d)^2+y^2=a^2}=\lambda/2\pi a\varepsilon$.

\section*{3. The atomic structure of PDADMA}\label{sec:atomPDA}

\begin{figure}
\begin{centering}
\includegraphics[width=1\linewidth]{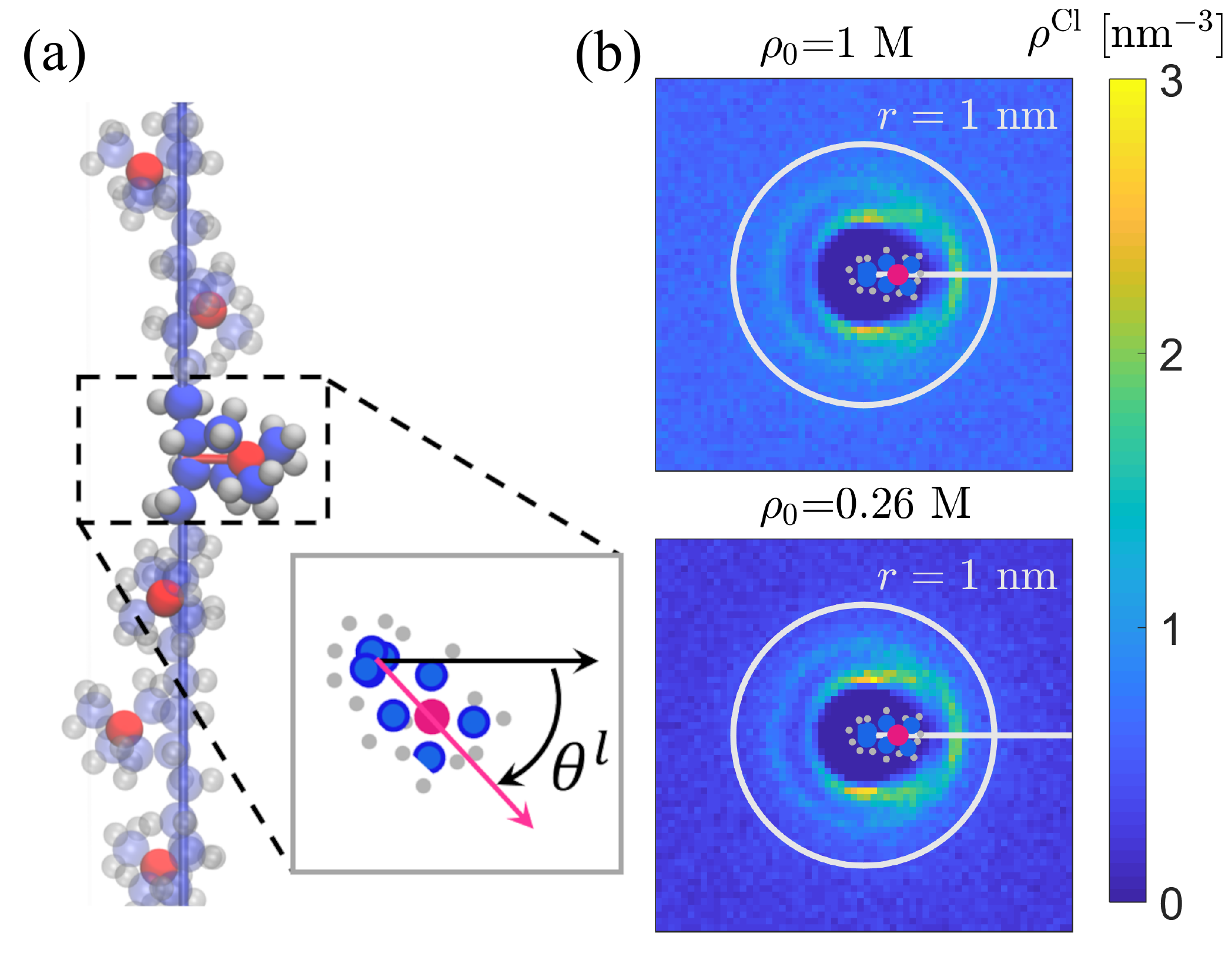}
\par\end{centering}
\caption{(a) The $l^{\rm th}$ monomer in the PDADMA is shown in the zoomed inset. The atomistic structure is projected onto the $(r,\theta)$ plane and the orientation of N$^{+}$ is recorded as $\theta^l$; the N$^{+}$, C and H atoms are colored in red, blue, and gray, respectively.
From (a) to (b), the orientation of a N$^{+}$ atom in the $l^{\rm th}$ monomer and the ion number density $\rho^{\textrm{Cl},l}$ around it are rotated by $-\theta^l$.
Then $\rho^{\textrm{Cl},l}$ are averaged over $l$ to get $\rho^{\textrm{Cl}}$ in panel (b) for different salt concentrations.
In the center of each density map we schematically show the atomistic structure of one monomer. 
\label{fig:densmap_PDA_mono}}
\end{figure}

Our atomistic PDADMA MD model consists of 10 monomers and the polymer axis is set along the $z$ direction. Each monomer corresponds to a backbone length of $L_z/10$.
Since the charged groups of the monomers adopt a helix-like structure (see Fig.~4(a) in the main text) for the straightened PDADMA chain, we define the location of N$^{+}$ of the $l^{\rm th}$ monomer by $(r^l, \theta^l, z^l), l=1, 2, ..., 10$ in cylindrical coordinates with longitudinal axis parallel to (and centered at) the backbone of the PE.
The ion number density around the $l^{\rm th}$ monomer (in the region $z^l-L_z/20< z < z^l+L_z/20$) is projected into a two-dimensional density map $\rho^{\textrm{Cl},l}(r,\theta)$ in polar coordinates.
The ion number densities around different monomers are similar but the angular orientation of the density follows the relative orientation of the monomer $\theta^l$. To account for this,
we rotate the ion number density map by $-\theta^{l}$ degrees as $\rho^{\textrm{Cl},l}(r,\theta- \theta^{l})$, to overlay the orientations of the monomers such that superposed ion number density data in terms of orientation with respect to monomer is obtained.
Then we take an average over $\rho^{\textrm{Cl},l}$ to get the density $\rho^{\textrm{Cl}}= \langle \rho^{\textrm{Cl},l}(r,\theta- \theta^{l})\rangle_l$ for different salt concentrations as shown in Fig.~\ref{fig:densmap_PDA_mono}(b).
This ion number density distribution around the PE monomers concretely shows the origins of the two-peak profile in the ion distribution in the main manuscript: ions form a semi-circular ring around the charged group with ion condensation localization visible both at the sides and at the tip of the charged group, leading to the binodal ion number density distribution. Similar findings were already reported in Ref.~\citenum{Batys2017}.
In particular, the Cl$^{-}$ ions stack mainly on the directions $\pm\pi/2$ and $0$, corresponding to the first and the second peaks of Cl$^-$ in the radial density profiles, respectively.
This configuration is caused by two main interactions on Cl$^-$ ions: attraction from positively charged group centered at N$^+$ atoms and steric repulsion from the surrounding neutral atoms. 
The positionally correlated configurations of the PE charge groups and its effect on the ion density span over different salt concentrations.

\section*{4. The orientational self-correlation function}\label{sec:corrPDA}

\begin{figure}
\begin{centering}
\includegraphics[width=0.9\linewidth]{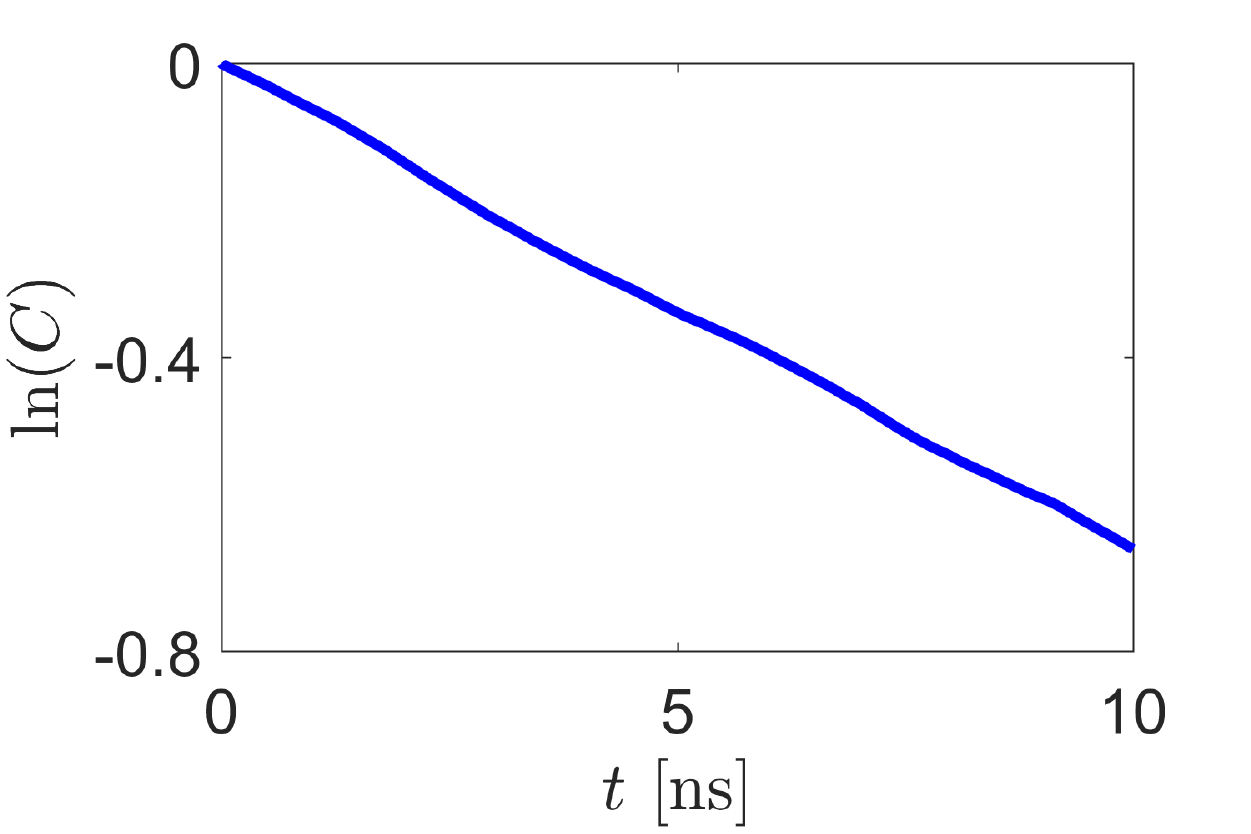}
\par\end{centering}
\caption{The orientational self-correlation function of a single PDADMA as a function of time $t$ on a semilogarithmic scale. The correlation time $t_{\rm c}=15$ ns corresponds to the slope of $\ln\left(C(t)\right)$.}
\label{fig:time_corr}
\end{figure}
The atomistic-detail PDADMA molecule spans the $z$ axial direction of the simulation box as a periodic molecule. However, it can both translate and rotate along the $z$ axis during the MD simulations. The orientation of the chain on the $z=0$ plane is given by the angle $\theta^1$ as defined in Fig.~\ref{fig:densmap_PDA_mono} ($l=1$).
During a 100~ns simulation we measured the time dependence of $\theta^1(t)$ and its time average $\bar \theta^1 = \langle \theta^1(t) \rangle_t$. We then define
the normalized orientational (fluctuation) self-correlation function $C(t)$ as
\begin{equation}
    C(t)= \frac{\langle \Delta \theta^1(s) \Delta \theta^1(s+t) \rangle_s}{\langle \Delta \theta^1(s) \Delta \theta^1(s) \rangle_s},
\end{equation}
where $\Delta \theta^1 = \theta^1 - \bar \theta^1$.
As an example, for a single PDADMA chain at salt concentration $\rho_{0}=0.26$ M, the self-correlation function decreases exponentially as $\exp{(-t/t_c)}$ as shown in Fig.~\ref{fig:time_corr}. From this data we can estimate the self-rotational correlation time as $t_{\rm c} \approx 15$ ns.
\section*{5. Comparison between SPB and SLPB theories}\label{sec:compSPB}

\begin{figure}[ht]
\begin{centering}
\includegraphics[width=0.9\linewidth]{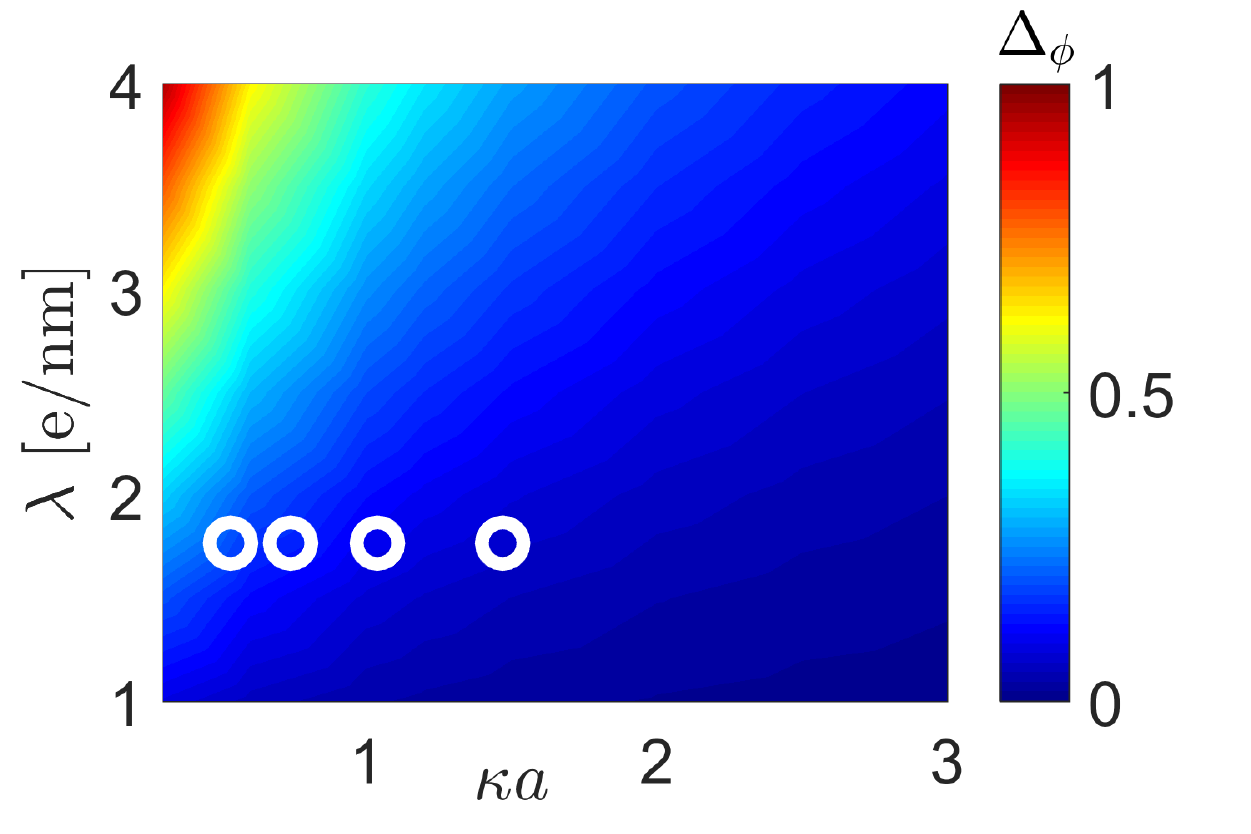}
\par\end{centering}
\caption{The mean relative error of the electrostatic potential $\Delta_{\phi}$ as a function of linear charge density $\lambda$ and dimensionless effective radius $\kappa a$.
The circles
represent the reference points for PDADMA
with concentrations $\rho_{0}=$ 0.13~M, 0.26~M, 0.52~M and 1.0~M, from left to right, respectively.
\label{fig:comp_PB_LPB}}
\end{figure}
\begin{figure}[ht]
\begin{centering}
\includegraphics[width=0.9\linewidth]{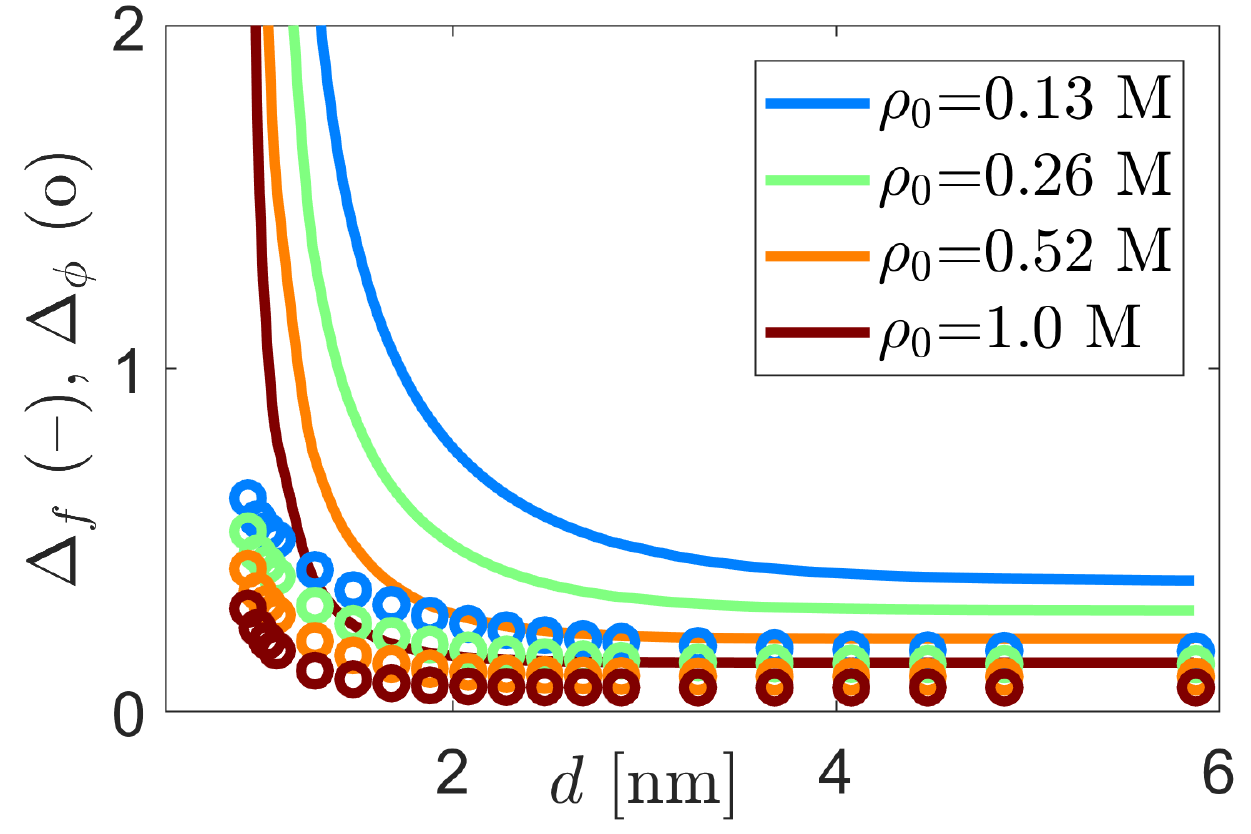}
\par\end{centering}
\caption{The relative errors in the mean force $\Delta_f$ (lines) and electrostatic potential $\Delta_{\phi}$ (circles) between the SPB and LSPB theories for the case of two PEs as a function of their separation.
\label{fig:comp_LPB_cyl2}}
\end{figure}
\begin{figure*}[t!]
\begin{centering}
\includegraphics[width=0.9\linewidth]{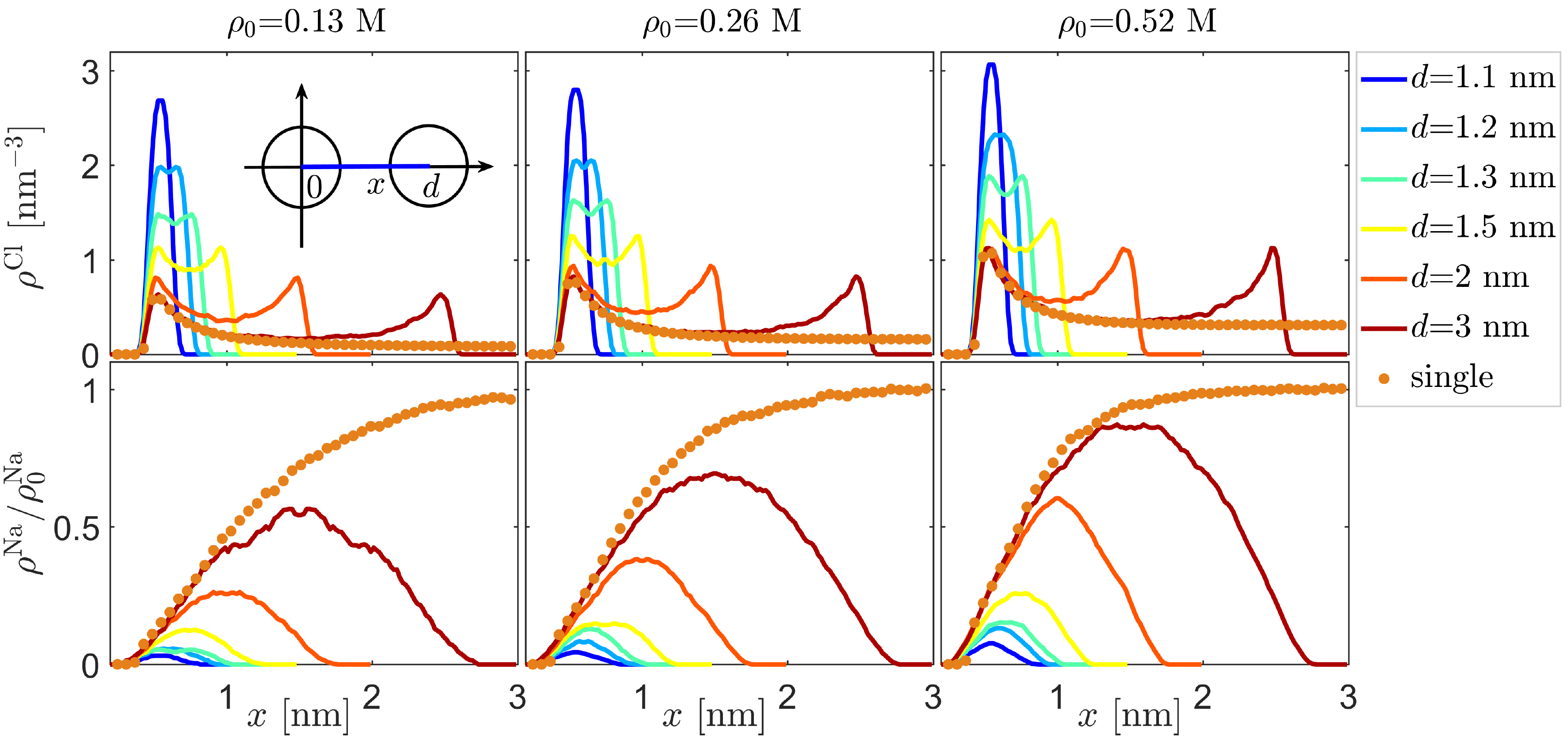}
\par\end{centering}
\caption{The Cl$^-$ and Na$^+$ number density profiles along the $x$ axis connecting the centers of the two PEs from the CG-MD model (as sketched in the inset).
The center-to-center distance $d$ changes from 1.1 nm to 3.0 nm.}
\label{fig:dens1D_2PDA}
\end{figure*}
\begin{figure}[b!]
\begin{centering}
\includegraphics[width=1.0\linewidth]{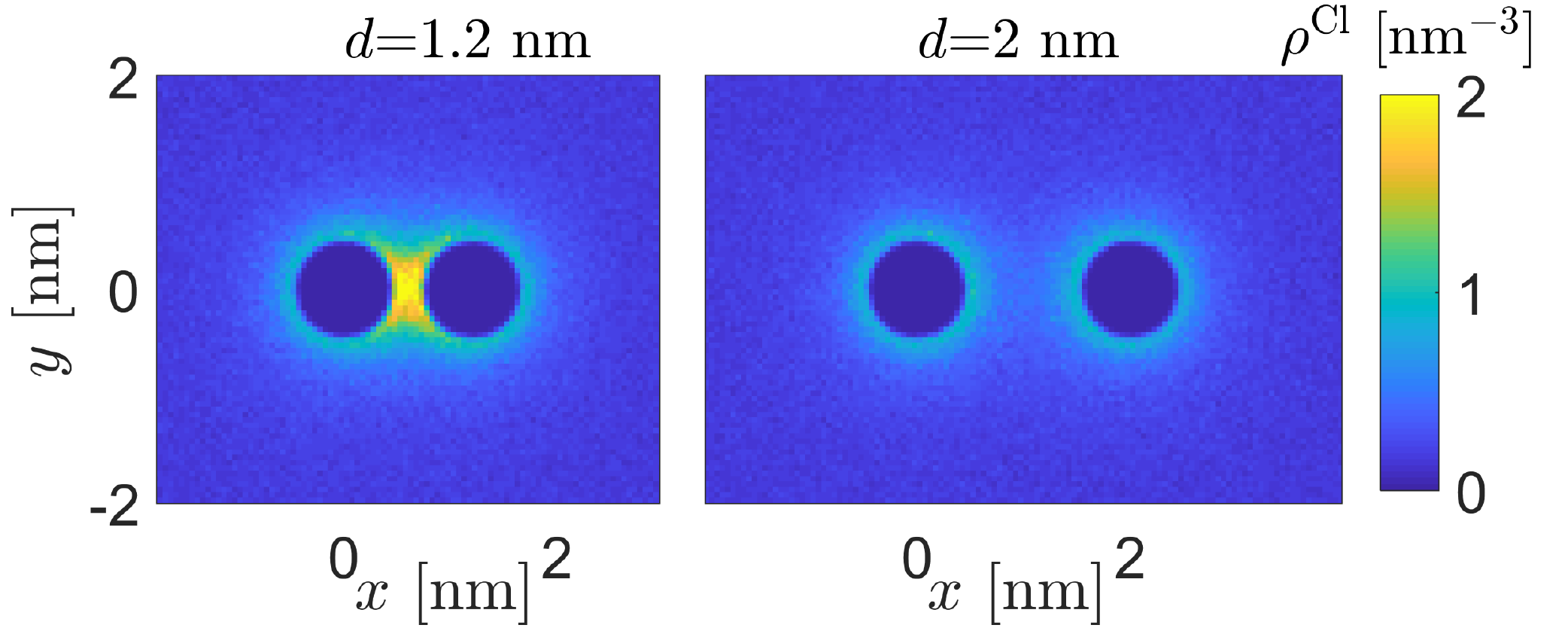}
\par\end{centering}
\caption{The Cl$^-$ density map around two rods with $d=1.2$ nm and $d=2.0$ nm from the CG-MD model. The added salt concentration is 0.26 M.}
\label{fig:density_2PDA}
\end{figure}
We addressed the differences in the electrostatic potential $\phi_{\rm SPB}(r)$ and $\phi_{\rm SLPB}(r)$ with the same linear charge density $\lambda$, effective radii $a_{\rm SLPB}=a_{\rm SPB}=a$ and monovalent salt concentration $\rho_{0}$.
A theoretical comparison between the two approximations is reported in Ref.~\citenum{Philip1970}. However, here we focus on PDADMA.
The difference is expressed as the relative error $\Delta_{\phi}\equiv\langle\left|\phi_{\textrm{SLPB}}(r)-\phi_{\textrm{SPB}}(r)\right|/\phi_{\textrm{SPB}}(r)\rangle_r$, which is averaged over $a<r<L_x/2$.
Using the Debye length $\kappa^{-1}$ as a unit of length, 
the only characteristic length scale in the system is $\kappa a$. Thus, $\Delta_{\phi}$ solely depends on $\lambda$ and $\kappa a$.
We show this in Fig.~\ref{fig:comp_PB_LPB}. As expected, for large values of $\lambda$, small values of $a$ and low salt concentration $\rho_{0}$ (corresponding to low $\kappa$), we find the highest-error region (red).
We can see that in the case of PDADMA (circles), the PB and LPB are almost identical at all salt concentrations considered here.
This is due to a relatively small linear charge density $\lambda=$ 1.767 e/nm, giving rise to small electrostatic potential $\phi$.
We conclude that in the case of PDADMA the linear approximation is successful in capturing the ion distribution.

For the two-PDADMA case, the relative error of mean force and electrostatic potential at $x=d/2$ are expressed as $\Delta_f\equiv\left|f_{\textrm{SLPB}}-f_{\textrm{SPB}}\right|/f_{\textrm{SPB}}$
and $\Delta_{\phi}\equiv\langle\left|\phi_{\textrm{SLPB}}(d/2,y)-\phi_{\textrm{SPB}}(d/2,y)\right|/\phi_{\textrm{SPB}}(d/2,y)\rangle_y$, respectively;
the latter is averaged over $-L_y/2< y < L_y/2$.
Their comparison is shown in Fig.~\ref{fig:comp_LPB_cyl2}.
In contrast to the single PE case, the mean force from SLPB is very different from that of the SPB, especially when the two PEs are close.

\section*{6. Ion density around two CG-MD PDADMAs}\label{sec:dens_PDA2}

In the CG model, we can examine the ion number densities around two PEs.
As shown in Fig.~\ref{fig:density_2PDA}, a significant fraction of the Cl$^-$ ions accumulate in the region between the two PEs when the distance between the two rods is relatively short ($d \leq 1.5$ nm). In Fig.~\ref{fig:density_2PDA} we report the two-dimensional number density for $d=1.2$ nm and $d=2.0$ nm.

To better quantify this, we obtained the one-dimensional cross section of the ion-number density profiles between the two PEs at $0< x <d, y=0$ (see the inset of Fig.~\ref{fig:dens1D_2PDA}) for different values of $d$, and compared them to the single-PE case, as shown in Fig.~\ref{fig:dens1D_2PDA}. When the two rods are close ($d=1.1$ nm), the peak value of the Cl$^-$ ion-number density is 3 to 4 times larger than that of the single-PE case. On the other hand, the Na$^+$ ion number density is much lower than that of a single rod. This is due to the stronger electrostatic repulsion, as well as the Cl$^-$ ions occupying more space in between the rods.

\section*{7. Limitation of the SPB approach}


In the original work of Vahid {\it et al.} \cite{Vahid2022} where the SPB theory was introduced it was applied to model monovalent ion distributions around single polystyrene sulphonate (PSS) molecules. The corresponding ion densities were accurately reproduced for a wide range of system parameters, including salt and ion sizes.
We performed additional testing for the case of two such PEs with atomistic-level MD simulations and the same setup as for the PDADMA molecules here. However, we found that there was an attractive force induced between two PSS molecules at short distances.
Such attraction is not expected in the weak-coupling regime and here it may be caused by entropy loss
due to the axially straight, infinite PE chain simulation setup.
Thus, we did not consider this case further in the present work.

\vspace{1cm}
\end{document}